\documentstyle[sprocl]{article}

\def\be{\begin{equation}}
\def\ee{\end{equation}}
\def\bea{\begin{eqnarray}}
\def\eea{\end{eqnarray}}

\begin{document}

\title{Probing Vacuum with Variational Methods in Quantum Gravity}
\author{Remo Garattini}

\address{M\'{e}canique et Gravitation, Universit\'{e} de Mons-Hainaut,
\\ Facult\'{e} des Sciences, 15 Avenue Maistriau,
\\ B-7000 Mons, Belgium\\ and\\ Facolt\`a di Ingegneria, Universit\`a degli Studi di
Bergamo,\\ Viale Marconi,5 24044 Dalmine (Bergamo), Italy
\\ E-mail: Garattini@mi.infn.it}

\maketitle\abstracts{We use variational methods to calculate quasilocal
energy quantum corrections. A comparison with the effective potential
calculated at quadratic order is made by means of gaussian wave functionals.
The method is a particular case of the effective action for composite
operators used in quantum field theory. In pure gravity the method is
applied for the first time. Implications on the foam-like scenario are
discussed.}

It is generally accepted that if Einstein gravity has a ground state, this
is represented by Flat space, at least at classical level. From the quantum
mechanical point of view, we know that at not zero temperature Flat space
decays via a nucleation of a black hole into a more stable space: the
Euclidean Schwarzschild space\cite{GPY}. Here we would like to discuss the
possibility of having a transition from flat space, at zero temperature, via
a neutral black hole pair creation\cite{Remo}. We shall investigate such a
possibility by means of a variational procedure applied to gaussian wave
functionals. Although it is not necessary for the forthcoming discussions,
let us consider the maximal analytic extension of the Schwarzschild metric,
i.e., the Kruskal manifold whose spatial slices $\Sigma $ represent
Einstein-Rosen bridges with wormhole topology $S^2\times R^1$. The complete
manifold ${\cal M}$ can be taken as a model for an eternal black hole
composed of two wedges ${\cal M}_{+}$ and ${\cal M}_{-}$ located in the
right and left sectors of a Kruskal diagram\cite{FroMar}. The hypersurface $%
\Sigma $ is divided in two parts $\Sigma _{+}$ and $\Sigma _{-}$ by a
bifurcation two-surface $S_0$. On $\Sigma $ we can write the gravitational
Hamiltonian 
\[
H_p=H-H_0=\int_\Sigma d^3x(N{\cal H+}N^i{\cal H}_i)
\]
\begin{equation}
+\frac 1\kappa \int_{S_{+}}^{}d^2xN\sqrt{\sigma }\left( k-k^0\right) -\frac
1\kappa \int_{S_{-}}d^2xN\sqrt{\sigma }\left( k-k^0\right) ,  \label{a1}
\end{equation}
where $\kappa =8\pi G$. The Hamiltonian has both volume and boundary
contributions. The volume part involves the Hamiltonian and momentum
constraints 
\begin{equation}
{\cal H}=\left( 2\kappa \right) G_{ijkl}\frac{\pi ^{ij}\pi ^{kl}}{\sqrt{^3g}}%
-\sqrt{^3g}R/\left( 2\kappa \right) =0,\quad {\cal H}_i=-2\pi _{i|j}^j=0,
\end{equation}
where $G_{ijkl}$ is the supermetric and $R$ denotes the scalar curvature of
the surface $\Sigma $. The volume part of the Hamiltonian $\left( \ref{a1}%
\right) $ is zero when the Hamiltonian and momentum constraints are imposed.
The Wheeler-DeWitt equation (WDW) is nothing but the Hamiltonian constraint
applied to a wave functional, i.e., ${\cal H}\tilde{\Psi}=0$. Nevertheless
we are interested in assigning a meaning to 
\begin{equation}
\frac{\langle \Psi \left| H_S-H_F\right| \Psi \rangle }{\langle \Psi |\Psi
\rangle }+\frac{\langle \Psi \left| H_{quasilocal}\right| \Psi \rangle }{%
\langle \Psi |\Psi \rangle },  \label{a2}
\end{equation}
where $\Psi $ is a wave functional whose structure will be determined later
and where $H_S$ $\left( H_F\right) $ is the total Hamiltonian referred to
the different spacetimes. Note that the first term of $\left( \ref{a2}%
\right) $ is simply the extension of the quasilocal energy subtraction
procedure generalized to the volume term. Note also that if the expectation
value is calculated on the wave functional solution of the WDW equation, we
obtain only the boundary contribution. Then to give meaning to $\left( \ref
{a2}\right) $, we adopt the semiclassical strategy of the WKB expansion. By
observing that the kinetic part of the Super Hamiltonian is quadratic in the
momenta, we expand the three-scalar curvature up to quadratic order. On the
other hand, following the usual WKB expansion, we will consider $\tilde{\Psi}%
\simeq C\exp \left( iS\right) $. In this context, the approximated WKB\ wave
functional will be substituted by a {\it trial wave functional} $\Psi \left[
g_{ij}\right] $ according to the variational approach we would like to
implement as regards this problem. To calculate the energy density, we need
that 
\begin{equation}
\frac{\left\langle \Psi \mid H_{|2}\mid \Psi \right\rangle }{\left\langle
\Psi \mid \Psi \right\rangle }=\frac{\int \left[ {\cal D}g_{ij}^{\bot
}\right] \int d_{}^3x\Psi ^{*}\left\{ g_{ij}^{\bot }\right\} {\cal H}%
_{|2}\Psi \left\{ g_{kl}^{\bot }\right\} }{\int \left[ {\cal D}g_{ij}^{\bot
}\right] \mid \Psi \left\{ g_{ij}^{\bot }\right\} \mid ^2}
\end{equation}
be stationary against arbitrary variations of $\Psi \left\{ h_{ij}\right\} $%
, where ${\cal H}_{|2}$ is the Hamiltonian density restricted to the TT
sector approximated up to quadratic order. Usually this procedure yields to
a Schr\"{o}dinger equation. Nevertheless to make calculations in practice,
we give to the trial wave functional the form 
\begin{equation}
\Psi \left[ h_{ij}^{\bot }\right] ={\cal N}\exp \left\{ -\frac
1{4l_p^2}\left\langle \left( g-\overline{g}\right) K^{-1}\left( g-\overline{g%
}\right) \right\rangle _{x,y}^{\bot }\right\} .  \label{a3}
\end{equation}
$K^{-1}\left( x,y\right) $ is the inverse {\it propagator, }parametrized as 
\begin{equation}
K^{\bot }\left( x,x\right) _{iakl}=\sum_N\frac{h_{ia}^{\bot }\left( 
\overrightarrow{x}\right) h_{kl}^{\bot }\left( \overrightarrow{y}\right) }{%
2\lambda _N\left( p\right) }
\end{equation}
where $\lambda _N\left( p\right) $ are infinite variational parameters and $%
h_{ia}^{\bot }\left( \overrightarrow{x}\right) $ are the eigenfunctions of
the differential operator 
\begin{equation}
\left( \triangle _2\right) _j^a:=-\triangle \delta _j^a+2R_j^a.  \label{a3a}
\end{equation}
$\triangle $ is the curved Laplacian and $R_j^a$ is the mixed Ricci tensor.
Thus $\left( \triangle _2\right) _j^a$ is nothing but the Lichnerowicz
operator projected on $\Sigma $. Choice $\left( \ref{a3}\right) $ is related
to the form of the Hamiltonian approximated to quadratic order in the metric
deformations. Indeed, up to this order we have a harmonic oscillator whose
ground state has a Gaussian form. Thus the one-loop-like Hamiltonian form
for TT deformations in both backgrounds, i.e. the Schwarzschild metric and
its reference metric, is\cite{Remo1} 
\begin{equation}
H^{\bot }=\frac 1{4l_p^2}\int_\Sigma ^{}d^3x\sqrt{g}G^{ijkl}\left[ K^{-1\bot
}\left( x,x\right) _{ijkl}+\left( \triangle _2\right) _j^aK^{\bot }\left(
x,x\right) _{iakl}\right] .
\end{equation}
Now we are in position to have the ({\it semiclassical}) explicit form of $%
\left( \ref{a2}\right) $. The classical term is given by 
\begin{equation}
\frac{\langle \Psi \left| H_{quasilocal}\right| \Psi \rangle }{\langle \Psi
|\Psi \rangle }=M-r\left[ 1-\left( 1-\frac{2MG}r\right) ^{\frac 12}\right] ,
\end{equation}
while the semiclassical part gives 
\begin{equation}
\frac{\langle \Psi \left| H_S-H_F\right| \Psi \rangle }{\langle \Psi |\Psi
\rangle }\simeq -\frac V{2\pi ^2}\left( \frac{3MG}{r_0^3}\right) ^2\frac
1{16}\ln \left( \frac{r_0^3\Lambda _{}^2}{3MG}\right) +a\frac i{MG},
\end{equation}
where $a$ is a positive number and $\Lambda $ is an UV cut-off. Apart the
imaginary contribution the previous result is independent of boundary
conditions since it is computed with respect to the volume term. Note that
the subtraction procedure at one loop has the correct ingredients to be
related to the Casimir energy. This seems to give some information about the
vacuum behaviour. Indeed, if we look at symmetric boundary conditions with
respect to the bifurcation surface $S_0$, then eq.$\left( \ref{a2}\right) $
can be interpreted as an energy gap measuring the possibility of creating a
black hole pair by a topological fluctuation; it also indicates that flat
spacetime is unstable with respect to pair creation. This seems to be an
instability of the zero temperature flat space with the total energy
conserved\cite{GPY,Witten,Remo}. Indeed the negative mode is a signal of the
passage from a false vacuum to the true one\cite{Coleman}. Note that this
calculation can be related to the quantity 
\begin{equation}
\Gamma _{{\rm 1-hole}}=\frac{P_{{\rm 1-hole}}}{P_{{\rm flat}}}\simeq \frac{%
P_{{\rm BlackHolePair}}}{P_{{\rm flat}}}.  \label{a4}
\end{equation}
To generalize a little more, suppose to enlarge this process from one pair
to a large but fixed number of such pairs, say $n$. What we obtain is a
multiply connected spacetime with $n$ holes inside the manifold, each of
them acting as a single bifurcation surface with the sole condition of
having symmetry with respect to the bifurcation surface even at finite
distance. Let us suppose the interaction between the holes can be neglected,
i.e., let us suppose that the total energy contribution is realized with a
coherent summation process. This is equivalent to saying that the wave
functional support (here, the semiclassical WDW functional) has a finite
size depending only on the number of the holes inside the spacetime. It is
clear that the number of such holes cannot be arbitrary, but is to be
related with a minimum size of Planck's order. Thus, assuming a coherency
property of the wave functional and therefore of the $N$-holes spacetime, eq.%
$\left( \ref{a4}\right) $ has to be generalized to 
\begin{equation}
\Gamma _{{\rm N-holes}}=\frac{P_{{\rm N-holes}}}{P_{{\rm flat}}}\simeq \frac{%
P_{{\rm foam}}}{P_{{\rm flat}}}.  \label{a5}
\end{equation}
Recall that a hole, here, has to be understood as a wormhole. Then, what eq.$%
\left( \ref{a5}\right) $ suggests is the possibility of generating a {\it %
foamy} spacetime\cite{Wheeler}, with wormholes as building blocks.

\section*{Acknowledgments}

I would like to thank Prof. A. Perdichizzi for his financial help and the
organizers who gave me the opportunity of presenting this talk.

\end{document}